\documentclass[preprint,12pt]{elsarticle}
\usepackage{amsmath,amssymb,amsfonts}
\usepackage{algorithmic}
\usepackage{graphicx}
\usepackage{textcomp}
\usepackage{xcolor}
\def\BibTeX{{\rm B\kern-.05em{\sc i\kern-.025em b}\kern-.08em
    T\kern-.1667em\lower.7ex\hbox{E}\kern-.125emX}}
\usepackage{tcolorbox}
\tcbuselibrary{listingsutf8}
\tcbset{
    myboxstyle/.style={
        colback=gray!10,
        colframe=gray!80,
        boxrule=0.5mm,
        arc=2mm,
        boxsep=5pt,
        left=5pt,
        right=5pt,
        top=5pt,
        bottom=5pt,
    }
}

\usepackage[colorlinks=true, allcolors=blue]{hyperref}
\usepackage{longtable}
\usepackage{threeparttable}
\usepackage{threeparttablex}
\usepackage{booktabs} 
\usepackage{array}     

\usepackage{appendix}
\usepackage{titlesec}

\usepackage[utf8]{inputenc}
\usepackage{xcolor}
\usepackage{listings}
\usepackage{geometry}
\usepackage{threeparttablex}
\usepackage{ragged2e}
\newcolumntype{L}[1]{>{\RaggedRight\arraybackslash}p{#1}}

\definecolor{vscode-bg}{RGB}{255,255,255}          
\definecolor{vscode-keyword}{RGB}{0,0,255}         
\definecolor{vscode-declaration}{RGB}{128,0,128}   
\definecolor{vscode-string}{RGB}{163,21,21}        
\definecolor{vscode-comment}{RGB}{0,128,0}        
\definecolor{vscode-ident}{RGB}{0,0,0}             
\definecolor{vscode-number}{RGB}{0,0,255}          
\definecolor{vscode-punctuation}{RGB}{0,0,0}       

\lstdefinelanguage{JavaScript}{
    keywords={break, case, catch, continue, do, else, false, for, function,
              if, in, new, null, return, switch, this, throw, true, try,
              typeof, void, while, with, instanceof},
    keywordstyle=\color{vscode-keyword}\bfseries,
    ndkeywords={const, let, var, class, export, extends, import, static,
                yield, await, async, super, delete},
    ndkeywordstyle=\color{vscode-declaration}\bfseries,
    identifierstyle=\color{vscode-ident},
    comment=[l]{//},
    morecomment=[s]{/*}{*/},
    commentstyle=\color{vscode-comment}\itshape,
    stringstyle=\color{vscode-string},
    morestring=[b]',
    morestring=[b]",
    sensitive=true,
}

\journal{Journal of Systems and Software}

\begin{document}
\begin{sloppy}
\begin{frontmatter}

\title{Do Comments and Expertise Still Matter? An Experiment on Programmers' Adoption of AI-Generated JavaScript Code}

\author[label1]{Changwen Li\corref{cor1}}
\ead{changwenl@student.unimelb.edu.au, arthurlicw@126.com}
\cortext[cor1]{Corresponding author.}

\author[label2]{Christoph Treude}
\ead{ctreude@smu.edu.sg}

\author[label1]{Ofir Turel}
\ead{oturel@unimelb.edu.au}

\affiliation[label1]{organization={The University of Melbourne},
            addressline={Grattan Street},
            city={Melbourne},
            postcode={3010},
            state={Victoria},
            country={Australia}}

\affiliation[label2]{organization={Singapore Management University},
            addressline={81 Victoria Street},
            city={Singapore},
            postcode={188065},
            state={Singapore},
            country={Singapore}}
    
\begin{abstract}
This paper investigates the factors influencing programmers' adoption of AI-generated JavaScript code recommendations within the context of lightweight, function-level programming tasks. It extends prior research by (1) utilizing objective (as opposed to the typically self-reported) measurements for programmers' adoption of AI-generated code and (2) examining whether AI-generated comments added to code recommendations and development expertise drive AI-generated code adoption. We tested these potential drivers in an online experiment with 173 programmers. Participants were asked to answer some questions to demonstrate their level of development expertise. Then, they were asked to solve a LeetCode problem without AI support. After attempting to solve the problem on their own, they received an AI-generated solution to assist them in refining their solutions. The solutions provided were manipulated to include or exclude AI-generated comments (a between-subjects factor). Programmers' adoption of AI-generated code was gauged by code similarity between AI-generated solutions and participants' submitted solutions, providing a behavioral measurement of code adoption behaviors. Our findings revealed that, within the context of function-level programming tasks, the presence of comments significantly influences programmers' adoption of AI-generated code regardless of the participants' development expertise.
\end{abstract}

\end{frontmatter}

\section{Introduction}
\label{sec:Introduction}
One of the most common hurdles in software engineering is turning ideas into programs \cite{3}. A code generation system capable of automatically translating natural languages into code could assist programmers, leading to a more efficient and productive programming paradigm \cite{1,2}. Automatically generating code from high-level descriptions has been a historic challenge for decades in the computer science community \cite{4}. The recent breakthrough in artificial intelligence, especially in large language models (LLMs), brings the computer science community closer to addressing this challenge \cite{1, 5}. AI has promising potential to revolutionize software engineering processes, including code search, code recommendation, automatic bug fixing, code review, risk assessment of code changes, and troubleshooting \cite{62}.

Despite the fact that recent advancements in artificial intelligence are reshaping software engineering \cite{5, 8}, little research has investigated the factors that influence programmers' adoption of AI-generated code \cite{89}. Ge \& Wu \cite{66} conducted an interview-based qualitative study to investigate the factors that impact the adoption of ChatGPT for bug fixing. They discovered several key influential factors, including performance expectancy, effort expectancy, social influence, facilitating conditions, data security, and trust \cite{66}. Russo conducted a survey-based study to investigate the factors influencing developers' adoption of generative AI tools in software engineering \cite{89}. Russo proposed the Human-AI Collaboration and Adaptation Framework (HACAF), a theoretical framework to explain programmers' adoption of AI coding tools \cite{89}. Liang et al. conducted a large-scale survey to evaluate the reasons programmers choose to use or avoid AI coding assistants by directly asking participants to answer why they use or refuse AI coding tools \cite{81}. 

While previous studies exploring the factors influencing programmers' adoption of AI-generated code have made valuable contributions, several research gaps remain. The first gap pertains to the over-reliance on self-reported data. The assessment of participants' adoption of AI was solely based on self-reported data, lacking complementary objective measurements. Previous studies used participants' self-reported intentions to gauge their adoption of AI-generated code or AI coding tools. For example, Russo used the question, "How likely are you to use a language model in your work in the next six months?" to measure participants' willingness to adopt AI coding tools \cite{89}. The reasons participants adopt or avoid AI-generated code were also collected through self-reported data by directly querying the reasons rather than conducting controlled experiments \cite{81, 89}. Some researchers \cite{164, 165, 166} have raised concerns about the limitations of self-reported questions in surveys. Self-reported data may not fully align with the actual usage pattern since self-reported data are subjective perceptions rather than objective facts. 

The second research gap is the non-comprehensive scope of the evaluated factors. Previous studies have primarily examined how productivity and programmers' perceptions influence the adoption of AI-generated code, neglecting some potentially crucial characteristics of programmers and the generated code. For instance, factors such as the presence of comments in AI-generated code and programmers' development expertise have been ignored.

To address the first research gap, we have designed a survey-based experiment employing objective measurements to assess participants' adoption of AI coding suggestions. In the survey, participants were required to complete a JavaScript programming task with the assistance of a code snippet generated by AI. This practical approach allows us to objectively assess their adoption of AI-generated code based on their modifications and usage of the provided snippet. Instead of merely asking participants to express their intentions to adopt AI-generated code, we analyze the submitted code to objectively determine adoption in controlled scenarios. We then conduct statistical analyses, treating potential influential factors as independent variables and the degree of adoption as the dependent variable. 

To bridge the second research gap, we undertook a systematic classification of the potential influential factors and evaluated the influence of some unexplored factors. We categorize the potential influential factors into two groups: (A) \textit{program factors} and (B) \textit{programmer factors}. Program factors refer to the factors inherent to the program itself, representing the characteristics of the AI-generated code. Programmer factors are tied to the attributes and conditions of the individual programmer. To keep the number of factors manageable in this study, we chose to investigate the presence of comments in the AI-generated code as the program factor and the programmers' development expertise as the programmer factor.


Our methodology not only deepens our understanding of the determinants of AI code adoption but also provides objective measurements of code adoption. Objective data is not necessarily superior to subjective self-reported data, but it could provide valuable complements to prior research.

\section{Related Work}
In this section, we introduce existing studies investigating factors that influence programmers' adoption of AI-generated code and AI coding tools. We also introduce the independent variables in this research: the presence of comments and development expertise, along with the control variable of attitude toward AI code generators. Since the influence of comments and development expertise on programmers' adoption of AI-generated code has not been studied before, we focus on introducing related work on their influence on code comprehension. 

\subsection{Factors Influencing Programmers' Adoption}
There is very little research examining the factors that influence programmers' adoption of AI code generators or AI-generated code. Previous related studies used interviews and surveys to investigate what impacts participants' adoption. Ge and Wu conducted semi-structured interviews with 50 participants to scrutinize factors influencing programmers' adoption of ChatGPT for bug fixing \cite{66}. The Unified Theory of Acceptance and Use of Technology (UTAUT) \cite{158}, a model used to explain users' intentions to adopt a technology, was adopted by Ge and Wu as their theoretical foundation \cite{66}. Their interviews included both open-ended and focused questions about how the four determinants in UTAUT, including performance expectancy, effort expectancy, social influence, and facilitating conditions, affect participants' adoption of AI-generated bug-fixing suggestions \cite{66}. Their study indicated that all of the four determinants in UTAUT, together with data security and programmers' trust, could influence the adoption of ChatGPT for bug fixing \cite{66}.

Russo conducted a survey-based study with 183 participants, using nine open-ended questions to investigate the factors influencing developers' adoption of generative AI tools in software engineering \cite{89}. Russo proposed the Human-AI Collaboration and Adaptation Framework (HACAF) to explain the adoption of generative AI tools in software engineering. HACAF is a theoretical model that combines insights from the Technology Acceptance Model (TAM) \cite{153}, Diffusion of Innovation (DOI) \cite{154}, and Social Cognitive Theory (SCT) \cite{155} to explain programmers' adoption of generative AI tools. HACAF emphasizes four determinant factors to programmers' adoption of AI coding tools, including perceptions about the technology, compatibility factors, social factors, and personal and environmental factors \cite{89}. Perceptions about the technology include perceived utility and ease of use. Compatibility factors measure the extent to which AI coding tools are compatible with current software engineering practices. Social factors include social influence and self-efficacy. Personal and environmental factors are personal willingness to experiment with AI coding tools and organizational support. 

Liang et al. also explored why programmers use or do not use AI code generators by conducting a survey-based study with 410 participants \cite{81}. Different from Russo's study \cite{89}, Liang et al. employed Likert-scale questions to assess the importance of various determinants that either motivate or discourage participants from using AI coding assistants \cite{81}. Their study indicates that the most significant reasons developers use AI code generators are the capability of AI tools to enhance productivity, including reducing keystrokes, accelerating programming, and assisting with syntax recalling. The pivotal reasons developers refuse to use AI coding tools are the inability of AI-generated code to fulfill specific requirements and the lack of control over producing desired outputs \cite{81}.

\subsection{Comments}
Although the impact of comments on the adoption of AI-generated code has not been systematically studied, related work on the impact of comments on code comprehension inspired our work. The significance of comments is emphasized in many coding guidelines because comments have a profound impact on code comprehension \cite{37}. Several empirical studies have explored the influence of comments on code comprehension \cite{38,39,40}. An early empirical study conducted by Takang et al. investigated whether the presence of comments influences code comprehension. Its findings demonstrated that comments play a profound role in enhancing program comprehension, regardless of the identifier naming convention \cite{38}. Nurvitadhi et al. expanded the study by examining how different types of comments influenced code comprehension \cite{39}. Nurvitadhi et al. investigated how class comments and method comments influence Java code understanding. Method comments notably enhance comprehension, while class comments do not have a significant impact \cite{39}. B{\"o}rstler and Paech further expanded the research by exploring how the quality of comments influences code comprehension \cite{40}. Their finding suggested that the mere presence of comments could enhance programmers' perception of the readability of the code regardless of the quality of the comments \cite{40}. 

\subsection{Development Expertise}
While previous studies have not thoroughly evaluated how development expertise influences programmers' adoption of AI-generated code, related research on its impact on code comprehension offers a valuable starting point. Development expertise comprises multiple aspects, including general programming experience, programming experience in particular languages, familiarity with task-specific knowledge, and personality traits \cite{116}. Previous studies have explored the influence of development expertise on code comprehension \cite{32, 45}. Lawrie et al. discovered that increased years of work experience and schooling resulted in a better capability for comprehending source code \cite{32}. Wagner and Wyrich studied the influence of intelligence and personality traits on code comprehension \cite{45}. The results of their study showed that there was a strong positive correlation between participants' general intelligence and their performance on code comprehension and a weaker negative correlation between participants' conscientiousness and code comprehension \cite{45}.

\subsection{Attitudes Towards AI Coding Tools}
Previous studies show that programmers have multifaceted attitudes toward AI code generators, including positive attitudes \cite{63, 65} and negative attitudes \cite{8, 63}. Several studies demonstrate that some programmers trust AI code generators and believe that AI code generators could improve their productivity \cite{63, 65}. Research has demonstrated that programmers adopt AI coding tools because they trust that AI coding tools could increase productivity \cite{81, 89}. On the other hand, besides positive attitudes, programmers hold negative attitudes as well \cite{8, 63}.  Feng et al. examined programmers' emotional responses to ChatGPT's code-generation capabilities by gathering social media posts regarding ChatGPT's code-generation features from Twitter and Reddit \cite{8}. Their analysis showed that the most common sentiments towards ChatGPT were fear, while happiness and anger were the two least common ones \cite{8}. Feng et al. concluded that programmers' fear derives from two sources, which are the opacity of ChatGPT and the sense of job threat \cite{8}. 

Kuhail et al. investigated programmers' fear of job loss \cite{63}. They investigated to what extent programmers fear job loss caused by AI code generators and why they fear or do not fear potential job loss. Participants in their study were required to rate their sense of current and future job security threats. The results suggest that only 24.2\% of participants consider AI code generators as non-trivial threats, categorizing them as high threats, moderate threats, or some threats. However, the outlook changes when considering the future, with approximately 50\% of the participants perceiving AI code generators as non-trivial threats to their job security. Kuhail et al. also discovered a positive correlation between programmers' trust in AI code generators and their fear of job replacements \cite{63}.

While prior research has investigated various factors influencing programmers' adoption of AI-generated code and AI coding tools, none have utilized objective measurements for participants' adoption of AI-generated code. In this work, we focus on how two previously unexplored factors -- AI-generated comments and developer expertise -- affect the adoption of AI-generated code. By conducting an experiment, we utilized objective measurements to gauge participants' adoption of AI-generated code and examine how these factors impact programmers' adoption behaviors, thereby addressing research gaps.

\section{Research Methodology}
\subsection{Research Questions}
In this study, we aim to evaluate the influence of two unexplored variables on programmers' adoption of AI-generated code: the presence of comments in the code (i.e., a program factor) and the programmers' development expertise (i.e., a programmer factor). We formulate the research questions \(\mathbf{RQ_1}\) and \(\mathbf{RQ_2}\) accordingly.

\begin{tcolorbox}[myboxstyle, boxsep=0pt, top=1pt, bottom=1pt]
\(\mathbf{RQ_1}\). How does ($A_1$) \textit{presence of comments} influence programmers’ adoption of AI-generated code?
\end{tcolorbox}

Our hypothesis was developed based on the Unified Theory of Acceptance and Use of Technology (UTAUT) \cite{158}. \hyperref[fig:utaut]{Fig. \ref{fig:utaut}} illustrates how the presence of comments shapes programmers’ adoption intentions and how developers’ expertise moderates the effect of comments on adoption intentions, with all relationships grounded in UTAUT. According to UTAUT, lower effort expectancy is linked to higher adoption intention \cite{158}. Moreover, existing research has underscored the critical role of comments in supporting code comprehension \cite{37}. Building on these theoretical and empirical foundations, we propose that the inclusion of comments in AI-generated code will reduce the time cost and cognitive burdens programmers incur to comprehend the code. This reduction aligns with UTAUT’s emphasis on effort expectancy, as lowered comprehension costs directly translate to decreased effort expectancy, ultimately influencing adoption intentions. Our formal hypothesis for \textbf{RQ$_1$} is: 

\noindent
\hangindent=2em \textbf{H$_{1}$}: The factor ($A_1$) \textit{presence of comments} increases programmers' adoption of AI-generated code. 

\begin{figure}[tb]
    \centering
    \includegraphics[width=0.75\columnwidth]{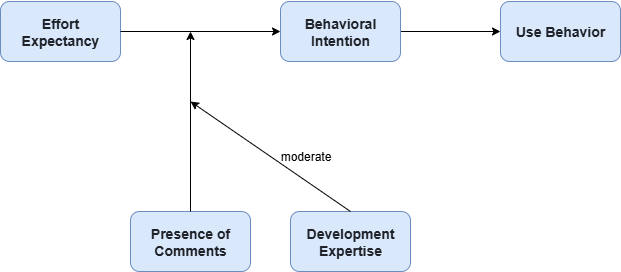}
    \caption{Hypothesis model based on UTAUT}
    \label{fig:utaut}
\end{figure}

\begin{tcolorbox}[myboxstyle, boxsep=0pt, top=1pt, bottom=1pt]
\(\mathbf{RQ_2}\). How does ($B_1$) \textit{development expertise} moderate the influence of ($A_1$) \textit{presence of comments} on programmers’ adoption of AI-generated code?
\end{tcolorbox}

Programmers with richer development expertise tend to be more experienced in programming and more familiar with task-specific knowledge \cite{116}, which could potentially enable them to have a better understanding of AI-generated code without comments. Aligning with UTAUT, experienced programmers—equipped with stronger code comprehension abilities—perceive a smaller reduction in effort expectancy when comments are present. This is because the reductions in time costs and cognitive burdens facilitated by comments hold less significance for them, given their capacity to effectively comprehend code regardless of whether comments are included. We hypothesize that the influence of ($A_1$) \textit{presence of comments} on programmers' adoption of AI-generated code would be moderated by ($B_1$) \textit{development expertise}, such that the positive effect of comments on adoption decreases as development expertise increases. The formal hypothesis for \textbf{RQ$_2$} is: 

\noindent
\hangindent=2em \textbf{H$_{2}$}: The positive effect of ($A_1$) \textit{presence of comments} on programmers' adoption of AI-generated code decreases as a function of ($B_1$) \textit{development expertise}.

\subsection{Experimental Design}
\subsubsection{Experiment Structure}

To ensure the reliability of our findings, we designed a survey-based experiment with a structured approach. By having participants first solve a programming problem on their own, we create a baseline for their understanding and approach. Then, by introducing an AI-generated solution with or without comments, we can objectively measure how these factors influence their adoption behaviors.

This methodology allows us to isolate the effects of comments and development expertise, providing a clear understanding of how these specific factors impact programmers' willingness to adopt AI-generated code.  Additionally, this design discourages participants from passively accepting AI-generated solutions, prompting them to critically evaluate and potentially integrate them with their own coding efforts. Conducting the experiment online with a consistent environment helps maintain control over external variables, ensuring that our results are robust and generalizable.

We developed a custom web application to host the survey on \url{https://survey.changwen-software-engineering.xyz}. We conducted pilot tests with five professional developers to ensure that there were no technical errors or usability issues in the survey. Participants in the pilot tests were recruited through the authors' professional network to ensure that they treated the study seriously and provided quality feedback. We conducted each run with a different participant and fixed the problems encountered in each run. 

The survey included six pages in total. The outline of the survey structure is presented in \hyperref[fig:survey-structure]{Fig. \ref{fig:survey-structure}}. The first page requested the participants' consent to participate in the study. Page one provided statements of consent forms, researchers' contact information, and a link to the plain language statement (PLS). After participants had opted to participate in the survey, the survey would navigate to the second page. 

\begin{figure}[tb]
    \centering
    \includegraphics[width=0.75\columnwidth]{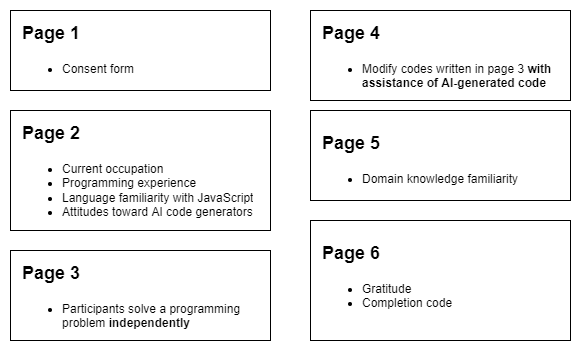}
    \caption{The outline of the survey structure}
    \label{fig:survey-structure}
\end{figure}

The second page collected participants’ demographics to capture the raw data for programmer factors, including ($B_{11}$) \textit{programming experience}, ($B_{12}$) \textit{programming language proficiency}, and ($B_{4}$) \textit{attitudes toward AI code generators}. We postponed measuring ($B_{13}$) \textit{domain knowledge familiarity} until page five to avoid biasing participants. Domain knowledge is the knowledge utilized to solve the programming problem on the third and fourth pages. Asking about domain knowledge upfront could unintentionally reveal what concepts were relevant to the upcoming programming task, influencing participants' responses. For example, if the task involved tree structures, asking about familiarity with trees in advance could hint at the solution and influence participants' responses.

On the third page, we showed participants the programming problems and required them to complete the programming problem in JavaScript \textbf{on their own}. We required participants to solve the programming problem independently first for two reasons. Firstly, it motivated participants to actively engage by trying to solve the problem themselves. In pilot tests, we observed that without this step, many participants would copy AI-generated solutions without fully comprehending the questions. Secondly, it helped us exclude potentially invalid responses, as participants unwilling to put in effort tended to quit the survey at this stage. This self-exclusion helped filter out less meaningful responses and improve data quality. We limited the programming language to JavaScript due to the limited participant number, ensuring manageable feasibility in the study.

The screenshot of the third page is shown in \hyperref[fig:code_editor]{Fig. \ref{fig:code_editor}}. As illustrated, the coding problem interface consists of several distinct regions, including the code editor region, test case region, timer region, and problem description region.

\begin{figure}[tb]
    \centering
    \includegraphics[width=\columnwidth]{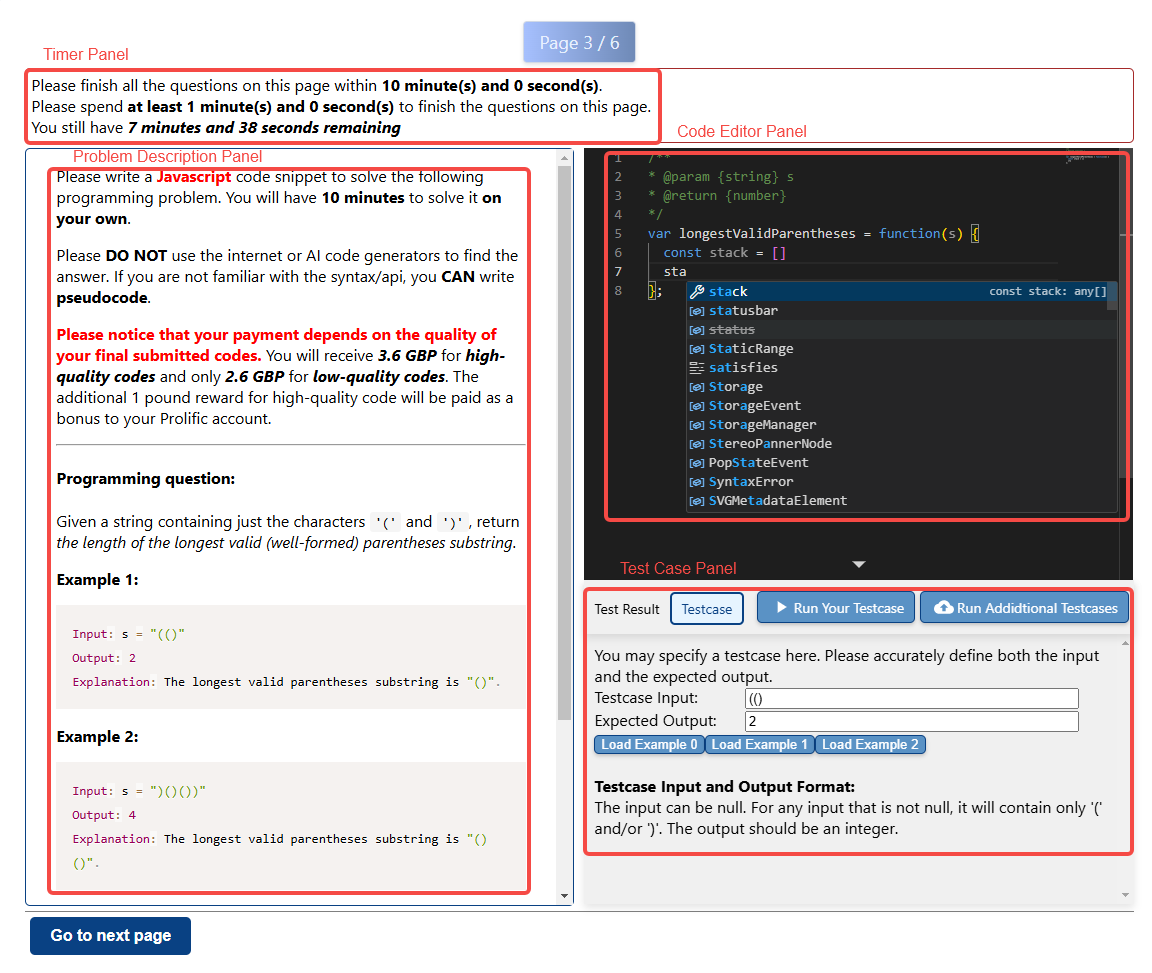}
    \caption{Overview of page three}
    \label{fig:code_editor}
\end{figure}

To provide participants with a coding experience similar to Microsoft Visual Studio Code (VSCode) \cite{163}, we integrated Monaco Editor, the code editor behind VSCode \cite{162}, into the code editor region. This integration enabled features such as syntax highlighting and auto-completion, allowing participants to quickly adapt to the environment.

In the test case region, participants could execute their code using either custom test cases or predefined ones. Each programming problem included ten predefined test cases, which were hidden from participants. As shown in \hyperref[fig:runtime_error]{Fig. \ref{fig:runtime_error}}, the test case region also supported switching to an output console, where participants could view comprehensive execution results, including error types, error messages, and the line number where an error occurred.

\begin{figure}[tb]
    \centering
    \includegraphics[width=0.75\columnwidth]{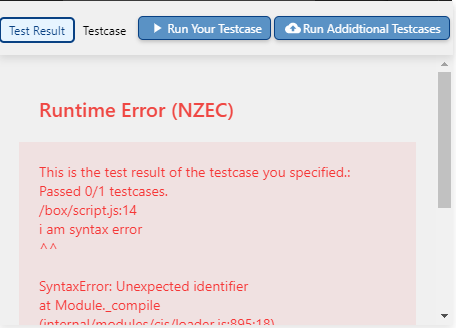}
    \caption{Overview of the test case region}
    \label{fig:runtime_error}
\end{figure}

A time limit was imposed on this page, ranging from a minimum of one minute to a maximum of ten minutes. The lower bound encouraged participants to make a genuine effort, while the upper bound simulated real-life time constraints and helped maintain a manageable study schedule.

On the fourth page, we provided a code snippet and notified the participants that it was generated by an AI code generator. They were also notified that they could modify the code written by themselves on page three with the assistance of the provided AI-generated code. Participants were explicitly informed that they had the option to use all the provided code, integrate part of the provided code into their original code, or choose not to use the provided code at all. We employed the same timer, code editor, and code execution features on page three. 

The fifth page of the survey gathered information on participants' familiarity with the domain knowledge (\textit{$B_{13}$}). 


On the sixth page, we expressed our gratitude to the participants. We provided a completion code for them and guided them to provide it so that we knew that they had finished the study. After receiving the completion code, we would pay them. 

\subsubsection{Motivation and Engagement Mechanisms}
\label{sec:Motivation and Engagement Mechanisms}
To motivate participants to focus on doing this survey and put in more effort, we have utilized two methods, including bonus payment and full-screen mode. For the bonus payment mechanism, we informed the participants that their reimbursement depended on the quality of the code they submitted on page three and page four. We told participants that they would only receive 2.6 GBP if the quality of their code was low and that they would receive 1 additional GBP as a bonus award if they submitted high-quality code. The 1 GBP payment increase was significant in this study because it was a 38\% increase in basic payment. In the end, we gave every participant both the basic 2.6 GBP reimbursement and the 1 GBP bonus regardless of the quality of their submitted code because it is hard to objectively determine the quality of their submitted code.

There were two purposes for adopting the bonus payment mechanism. The first purpose was to ensure genuine engagement from participants. Performance-based incentives could help increase participants’ performance \cite{85}. The difference in payment could motivate participants to put more effort into their participation. The second purpose was to motivate participants to provide more representative data. Programmers are often required to write high-quality code in their daily work. Therefore, motivating participants to produce high-quality code in our survey was a way to motivate them to approach the task as if they were coding in their actual professional work. 

The second motivation mechanism was full-screen mode, which encouraged participants to stay in full-screen mode while doing the survey. We motivated participants to stay in full-screen mode for two reasons. Firstly, staying in full-screen mode helps participants focus on the survey. Secondly, maintaining the survey in full-screen mode helps to prevent participants from seeking external assistance, such as information from the Internet and solutions created by the participants themselves using AI tools. Our goal was that participants only interact with the AI-generated solutions provided in our survey to gauge their adoption accurately. External information could influence their adoption of the provided AI-generated code in the survey, which might introduce bias.


We used a dialogue window to guide participants to stay in full-screen mode. After participants signed the consent form, they would see a dialogue window as shown in \hyperref[fig:full-screen1]{Fig. \ref{fig:full-screen1}}.  This dialogue window informed the participants that they needed to stay in full-screen mode, and their behavior of leaving full-screen mode would be recorded. We also warned them that leaving full-screen mode may lead to no reimbursement, although we still paid them even if they had left full-screen mode. We recorded this data to enable us to discard participant data that may have been tainted by seeking outside assistance. Once participants clicked the OK button, their browsers were switched to full-screen mode. Once they left full-screen mode, this dialogue window shown in \hyperref[fig:full-screen1]{Fig. \ref{fig:full-screen1}} would appear again. This survey design was approved by the ethics review board of our institution and follows the rules of the Prolific data collection platform used in this work.

\begin{figure}[tb]
    \centerline{\includegraphics[width=0.75\columnwidth]{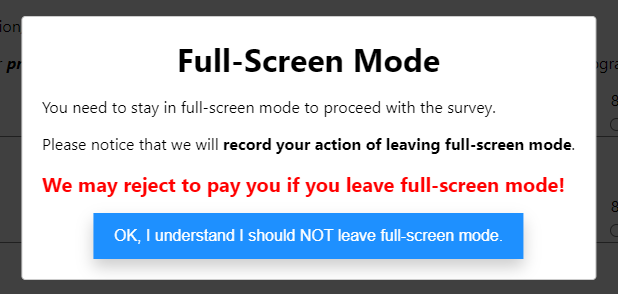}}
    \caption{Full-screen mode dialogue window while not in full-screen mode}
    \label{fig:full-screen1}
\end{figure}

\subsection{Experimental Objects}
The experimental objects contained programming problems, the corresponding predefined test cases, and AI-generated solutions to the programming problems. 

\subsubsection{Programming Problems}
We have chosen five programming problems from LeetCode, a well-known and popular platform where programmers practice programming \cite{49, 82}. Each survey will be randomly assigned one of these five programming problems and the participant needs to solve this programming problem on the third page and the fourth page of the survey. We chose programming problems from LeetCode for the following two reasons. Firstly, LeetCode's popularity increases the chances of high-quality and correct programming problems. The programming problems from LeetCode have clear introductions and well-crafted content. Moreover, LeetCode programming problems have been proven to be good experimental materials, as they have been used in multiple studies to evaluate AI code generation \cite{63, 64, 80}. 

To enhance the generalizability of this study, we chose five programming problems with different topics and different difficulty levels. According to the topic tags provided by LeetCode, the topics of these programming problems include map, stack, queue, linked list, and tree. We have chosen two easy programming problems, two medium programming problems, and one hard programming problem based on the difficulty labeled by LeetCode. A brief summary of the five selected programming problems is presented in \hyperref[table:problem-summary]{Table \ref{table:problem-summary}}.

\begin{table}[htb]
\centering
\begin{threeparttable}
\caption{A brief summary of selected programming problems}
\label{table:problem-summary}
\begin{tabular}{p{0.25\columnwidth} p{0.15\columnwidth} p{0.15\columnwidth} p{0.15\columnwidth}}
\toprule 
Name & Topics & Difficulty Level & LeetCode Number\tnote{*} \\
\midrule 
Ransom Note & Map & Easy & 383 \\
Remove Duplicates from Sorted List & Linked List & Easy & 83 \\
Find the Winner of the Circular Game & Queue & Medium & 1823 \\
Validate Binary Search Tree & Tree & Medium & 98 \\
Longest Valid Parentheses & Stack & Hard & 32 \\
\bottomrule 
\end{tabular}
\begin{tablenotes}
\item[*] LeetCode Number is a unique identifier assigned to each problem on the LeetCode platform for easy reference and search.
\end{tablenotes}
\end{threeparttable}
\end{table}

\subsubsection{Predefined Test Cases}
Participants were able to execute their code using predefined test cases on both page three and page four of the survey. Each programming problem included 10 pre-defined test cases, all of which were hidden from participants. These test cases were generated by GPT 4 and subsequently reviewed by the authors of this paper to ensure their correctness and scalability.

\subsubsection{AI-generated Solutions}
We have created two versions of AI-generated solutions for each programming problem: solutions \textbf{with} comments and solutions \textbf{without} comments. Every survey was assigned one of these two versions. On the fourth page of the survey, participants used the assigned AI-generated solutions to assist in modifying the code written by themselves on page three.

The provided AI-generated code snippets were generated by ChatGPT 4 Turbo. To generate the solutions, we copied the problem description, including examples, into ChatGPT 4 Turbo. Then, we added the prompt "Please solve this problem in JavaScript" and included the template code from LeetCode, which provides function signatures. We then added a final prompt based on the version: "Please do not add any comment" for the no-comment version and "Please add meaningful comments" for the commented version. We confirmed that all the code snippets generated by ChatGPT in this research could be executed correctly and passed all test cases on LeetCode. We also verified that all comments in the AI-generated code were meaningful and contributed to understanding the solution, rather than being irrelevant or nonsensical. An example of the AI-generated solution with and without comments is shown in \hyperref[fig:comments]{Fig. \ref{fig:comments}}.

\begin{figure}[htbp]
    \begin{minipage}[t]{0.48\textwidth}
        \begin{lstlisting}[caption=An example of the AI-generated code without comments, label=lst:iterative]
/**
 * @param {string} s
 * @return {number}
 */
var longestValidParentheses = function (s) {
  let maxLen = 0;
  let stack = [-1];

  for (let i = 0; i < s.length; i++) {
    if (s[i] === "(") {
      stack.push(i);
    } else {
      stack.pop();
      if (stack.length === 0) {
        stack.push(i);
      } else {
        const currentLen = i - stack[stack.length - 1];
        maxLen = Math.max(maxLen, currentLen);
      }
    }
  }

  return maxLen;
};
        \end{lstlisting}
    \end{minipage}
    \hfill
    \begin{minipage}[t]{0.48\textwidth}
        \begin{lstlisting}[caption=An example of the AI-generated code with comments, label=lst:recursive]
/**
 * @param {string} s
 * @return {number}
 */
var longestValidParentheses = function (s) {
  let maxLen = 0;
  // Use a stack to track the indices of unmatched parentheses.
  // -1 handles edge cases where the substring starts at index 0
  let stack = [-1];

  for (let i = 0; i < s.length; i++) {
    if (s[i] === "(") {
      stack.push(i);
    } else {
      // Pop for matching ')'
      stack.pop();
      if (stack.length === 0) {
        // If stack is empty, it means this ')' is unmatched. 
        // Push its index to reset the base for future valid substrings.
        stack.push(i);
      } else {
        // Calculate current valid substring length
        const currentLen = i - stack[stack.length - 1];
        maxLen = Math.max(maxLen, currentLen);
      }
    }
  }

  return maxLen;
};
        \end{lstlisting}
    \end{minipage}
    \caption{An example of the AI-generated code with and without comments}
    \label{fig:comments}
\end{figure}

\subsection{Variables, Measurements, and Analyses}
\hyperref[tab:independent-variables]{Table \ref{tab:independent-variables}} demonstrates the variables in this study, including independent variables, control variables, and dependent variables. 

\begin{table}[htbp]
\centering
\begin{threeparttable}
\caption{Outline of variables}
\label{tab:independent-variables}
\begin{tabular}{p{0.15\columnwidth} p{0.4\columnwidth} p{0.3\columnwidth}}
\toprule
Abbreviation & Full Name & Category \\
\midrule
$A_{1}$ & Presence of comments\tnote{a} & Independent variable \\
$B_{1}$ & Development expertise\tnote{b} & Independent variable \\
$B_{11a}$ & Perceptions of programming experience & Independent variable \\
$B_{11b}$ & Years of overall programming experience & Independent variable \\
$B_{12a}$ & Perceptions of programming language proficiency & Independent variable \\
$B_{12b}$ & Years of experience in JavaScript development & Independent variable \\
$B_{13a}$ & Perceptions of domain knowledge familiarity & Independent variable \\
$B_{13b}$ & Prior exposure to domain knowledge & Independent variable \\
$B_{2}$ & Gender\tnote{c} & Control variable \\
$B_{3}$ & Age\tnote{d} & Control variable \\
$B_{4}$ & Attitudes toward AI code generators\tnote{e} &  Control variable \\
$B_{4f}$ & Fear of AI code generators & Control variable \\
$B_{4a}$ & Attitudinal acceptance of AI code generators & Control variable \\
$Y$ & Latent code similarity\tnote{f} & Dependent variable \\
\bottomrule
\end{tabular}
\begin{tablenotes}
\item[a] The variable "presence of comments" is manipulated by providing AI-generated code snippets either with comments or without them.
\item[b] The variable "development expertise" is obtained through PCA analysis on $B_{11a}$, $B_{11b}$, $B_{12a}$, $B_{12b}$, $B_{13a}$, and $B_{13b}$.
\item[c] Data of the variable "gender" is provided by Prolific. 
\item[d] Data of the variable "age" is provided by Prolific. 
\item[e] Attitudes toward AI code generators comprise two aspects, which are fear and attitudinal acceptance. 
\item[f] Latent code similarity is computed through PCA analysis on four different code similarity metrics in \hyperref[tab:similarity-metrics]{Table \ref{tab:similarity-metrics}}.

\end{tablenotes}
\end{threeparttable}
\end{table} 

\subsubsection{Independent Variables}
The independent variables in this research were ($A_1$) \textit{the presence of comments} and ($B_1$) \textit{development expertise}. The variable $A_1$ was a categorical variable determined by whether the AI-generated code included comments. The value of $A_1$ was 1 if comments were present, and the value of $A_1$ was 0 if comments were absent.

To capture the complexity of ($B_1$) \textit{development expertise}, the corresponding factor $B_1$ has three sub-factors, including ($B_{11}$) \textit{programming experience}, ($B_{12}$) \textit{programming language proficiency}, and ($B_{13}$) \textit{domain knowledge familiarity}. We assessed ($B_1$) \textit{development expertise} by performing PCA (Principal Component Analysis) on $B_{11}$, $B_{12}$, and $B_{13}$. The detailed PCA analysis will be introduced in \hyperref[sec:Development Expertise]{Section \ref{sec:Development Expertise}}.

To enhance the reliability of the obtained data, we utilized two questions to assess each sub-factor. First, we employed a Likert scale question ($B_{11a}$, $B_{12a}$, $B_{13a}$) for each sub-factor, translating participants' perceptions into integers from 1 (Not experienced/proficient/familiar at all) to 5 (Extremely experienced/proficient/familiar). For sub-factors $B_{11}$ and $B_{12}$, we used questions $B_{11b}$ and $B_{12b}$ to gauge participants' years of experience in overall programming and JavaScript development, respectively. The second question used to assess $B_{13}$, question $B_{13b}$, measured participants' familiarity with specific domain knowledge by determining their level of prior exposure to certain topics. The possible levels were "I have never heard of \textit{x}," "I have heard of \textit{x} but never used it," "I have occasionally used \textit{x}," and "I have regularly used \textit{x}" (here, \textit{x} represents a certain domain knowledge topic). 

\subsubsection{Control Variables}
The control variables are ($B_2$) \textit{gender}, ($B_{3}$) \textit{age}, and ($B_{4}$) \textit{attitudes towards AI Code Generators}. We evaluated ($B_{4}$) \textit{attitudes towards AI Code Generators} by mirroring and modifying the Attitudes Towards Artificial Intelligence (ATAI) scale proposed by Sindermann et al. \cite{51}.  Akin to the ATAI scale, we categorized participants' attitudes toward AI code generators into two domains: fear and attitudinal acceptance. Therefore, we had two sub-factors for \textit{$B_4$}, which are ($B_{4f}$) \textit{fear of AI code generators} and ($B_{4a}$) \textit{attitudinal acceptance of AI code generators}. We explored $B_4$ through grid Likert scales in \hyperref[fig:grid_likert]{Fig. \ref{fig:grid_likert}}, where the mean value of the responses to the first two questions represents $B_{4a}$, and the mean value of the responses to the last three questions represents $B_{4f}$. 

\begin{figure}[tb]
    \centering
    \includegraphics[width=0.9\columnwidth]{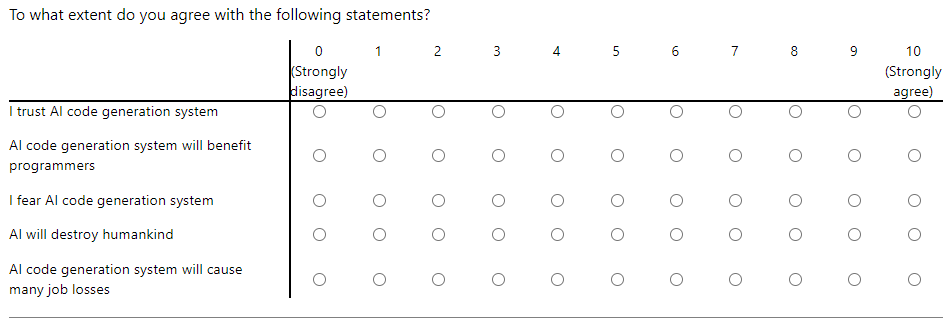}
    \caption{Grid Likert-scale question on evaluating participants' attitudes towards AI code generators \cite{51}}
    \label{fig:grid_likert}
\end{figure}

\subsubsection{Dependent Variable}
\label{subsec:Dependent Variable}
Building on research on algorithm appreciation \cite{167}, which suggests that individuals often rely on algorithmic outputs even without consciously acknowledging their influence, we define code adoption as \textbf{behavioral reliance on AI-generated code} rather than verbal declaration. Consistent with prior research that favors observable behavior \cite{167}, we measure this reliance using code similarity metrics—that is, the extent to which participants incorporated AI-generated code into their final submissions.

To operationalize this definition, we devised a novel objective metric. Participants were tasked with writing a code snippet with the assistance of an AI-generated solution (page four of the survey). The degree of adoption was then quantified by comparing the similarity between the code submitted by the participants and the AI-generated code provided on page four. This method provides an objective assessment of the code adoption behavior.


To gain insights into the dependent variable from different dimensions, four distinct code similarity metrics were utilized in this study, including line-level code similarity from the participant perspective (\textit{LP}), line-level code similarity from the AI perspective (\textit{LA}), token-sequence-level code similarity from the participant perspective (\textit{TP}), and token-sequence-level code similarity from the AI perspective (\textit{TA}). These metrics could be categorized by two dimensions: the dimension of entity perspective and the dimension of code similarity level. \hyperref[tab:similarity-metrics]{Table \ref{tab:similarity-metrics}} shows the classification and interpretation of these metrics. For data analysis, we created a new latent variable, \textit{latent code similarity}, by combining \textit{LP}, \textit{LA}, \textit{TP}, and \textit{TA} through PCA. The detailed PCA analysis is presented in \hyperref[sec:Latent Code Similarity]{Section \ref{sec:Latent Code Similarity}}. The new latent variable, \textit{latent code similarity}, was the dependent variable for hypotheses testing. 

The dimension of entity perspective indicated the entity from which the similarity metric was assessed, suggesting which entity was used as the denominator when calculating similarity. The possible values for the entity perspective dimension were participant and AI. If the entity perspective of a code similarity metric is "participant," then the metric represents the percentage of the participant's code that is similar to the AI-generated code, highlighting to what extent the participant's coding approach has been influenced by the AI-generated solution. If the entity perspective of a code similarity metric is "AI," then the metric represents the percentage of the AI-generated code similar to the participant's code, revealing the percentage of AI-generated code used by the programmer. 

The dimension of code similarity level showed the level at which the comparison was conducted. The possible values for the dimension of code similarity level were line level and token sequence level. The line-level code similarity metrics calculated the ratio of identical lines to total lines. Only lines that were exactly the same in both the AI-generated code and the participants' submitted code would be counted as a match (differences in whitespace and comments were ignored). Token-sequence-level code similarity metrics compute similarity by comparing the token sequences of two code snippets, a process carried out using Jplag \cite{101}. The token is a basic element of the code abstraction \cite{101, 102}. A token sequence represents some fractions of the code snippet without necessarily representing a line of code. Token sequence abstracts the code, providing a mapping to the code structure without being influenced by different identifier names and comments \cite{102}. 

Some code similarity-checking tools were developed to detect plagiarism in students' programming assignments at a token-sequence level \cite{53}, including MOSS \cite{102} and JPlag \cite{101}. We used JPlag \cite{101} to calculate the code similarity at the token-sequence level\footnote{We utilized the JPlag Java package from \url{https://github.com/jplag/JPlag} to calculate the token-sequence-level similarity}. Line-level code similarity investigated participants' direct copy-and-paste behavior. In contrast, the token-sequence-level code similarity focused on the abstract structure of the code.

JPlag captures code similarity at an abstract, structural level by detecting patterns such as iteration loops, control flow, and the use of data structures in the AI-generated code \cite{101}. This enables JPlag to identify participants’ adoption of AI-generated algorithms even when the code is not copied verbatim but reimplemented with similar underlying logic. For example, in the case shown in \hyperref[fig:comparison]{Fig. \ref{fig:comparison}}, both the AI-generated code and the participant’s submission used recursion to validate a binary tree. Despite this shared algorithmic structure, the two code snippets appeared quite different on the surface, with only two exact line matches. The line-level similarity from the AI perspective (LA) was just 0.13, whereas the token-sequence-level similarity from the AI perspective (TA), as computed by JPlag, was 0.75. This example illustrates JPlag’s ability to capture structural similarity beyond superficial textual resemblance.

\begin{figure}[htbp]
    \begin{minipage}[t]{0.48\textwidth}
        \begin{lstlisting}[caption=An example of the AI-generated code, label=lst:iterative]
var isValidBST = function (root) {
  function validate(node, min, max) {
    // An empty tree is a valid BST
    if (!node) return true;

    // The node's value must be within the min and max bounds
    if (
      (min !== null && node.val <= min) ||
      (max !== null && node.val >= max)
    ) {
      return false;
    }

    // Recursively check the left subtree and right subtree
    // For the left subtree, update the max bound
    // For the right subtree, update the min bound
    return (
      validate(node.left, min, node.val) && validate(node.right, node.val, max)
    );
  }

  // Initially, there are no bounds on the root's value
  return validate(root, null, null);
};
        \end{lstlisting}
    \end{minipage}
    \hfill
    \begin{minipage}[t]{0.48\textwidth}
        \begin{lstlisting}[caption=An example of the user-submitted code, label=lst:recursive]
var isValidBST = function (root) {
  return isValidNode(root, -Infinity, Infinity);
};

function isValidNode(node, min, max) {
  if (node === null) return true;

  const curValidate = node.val > min && node.val < max;

  // recursively check left subtree
  const leftValidate = isValidNode(node.left, min, node.val);

  // recursively check right subtree
  const rightValidate = isValidNode(node.right, node.val, max);

  return curValidate && leftValidate && rightValidate;
}
        \end{lstlisting}
    \end{minipage}
    \caption{A comparison between AI-generated code and user-submitted code}
    \label{fig:comparison}
\end{figure}


\begin{longtable}{p{0.15\columnwidth} p{0.15\columnwidth} p{0.15\columnwidth} p{0.35\columnwidth}}
    \caption{Classification of code similarity metrics}
    \label{tab:similarity-metrics}\\
    \toprule
    Abbreviation & Entity Perspective \newline Dimension & Code Similarity Level Dimension & Interpretation \\
    \midrule
    \endfirsthead
    \caption[]{Classification of code similarity metrics (continued)}\\
    \toprule
    Abbreviation & Entity Perspective \newline Dimension & Code Similarity Level Dimension & Interpretation \\
    \midrule
    \endhead
    \bottomrule
    \endfoot
    \bottomrule
    \endlastfoot
    LP & Participant & Line level & Percentage of the participant's code that are similar to the AI-generated code at the line level. \\
    LA & AI & Line level & Percentage of the AI-generated code that is similar to the participant's code at the line level. \\
    TP & Participant & Token sequence level & Percentage of the participant's code that are similar to the AI-generated code at the token sequence level. \\
    TA & AI & Token sequence level & Percentage of the AI-generated code that are similar to the participant's code at the token sequence level. \\
\end{longtable}

\subsection{Participants}
Participants in this study were recruited from Prolific \cite{95}, an online crowd-sourcing academic data collection platform used extensively in multiple research fields \cite{95}, including software engineering \cite{89, 96, 97}. This study strictly followed the applicable guidelines for ethical conduct in human research and was approved by the ethics review board of our institution.


\subsubsection{Pre-Screening Mechanism}
\label{sec:Pre-Screening Mechanism}
Participants were pre-screened to ensure they met the study's eligibility criteria. We utilized pre-screeners provided by Prolific to select participants with programming experience in JavaScript to ensure they were capable of doing the programming tasks in the survey. To minimize misunderstanding of the English instructions in our survey, we further pre-screened participants to ensure their primary language was English. The last pre-screener excluded participants with approval rates below 95\%. The approval rate indicates the percentage of submissions approved by researchers. A low approval rate could indicate a participant's history of negative behaviors, such as submitting low-effort responses \cite{99}. To mitigate low-effort responses in our survey, we required participants to have at least a 95\% approval rate.

\subsubsection{Post-Survey Screening Mechanism}
\label{sec:Post-Survey Screening Mechanism}
We employed two post-survey screening approaches to filter out invalid responses. The first approach was excluding responses that might have sought external assistance. If participants left full-screen mode more than twice or for longer than two seconds, then they were deemed as having sought external assistance. As illustrated in \hyperref[sec:Motivation and Engagement Mechanisms]{Section \ref{sec:Motivation and Engagement Mechanisms}}, using external resources could introduce bias to participants' adoption of the provided AI-generated code. Therefore, these responses were viewed as invalid responses and excluded from the data analysis.

The second post-survey screening approach was to exclude low-effort responses. A low-effort response has a damaging impact on survey-based research as it introduces untrustworthy data \cite{100}. To examine the level of effort of a response, we manually reviewed the code participants submitted on page 4. We examined whether the complexity of the submitted code was significantly low. For instance, when required to validate whether a tree is a valid binary search tree, a participant submitted a solution that directly returned true. This participant returned a hard-coded value rather than trying to solve the programming problem. An overly simplistic code that did not address the programming problem suggested low effort. 

In this study, we collected responses from 251 participants in total. After employing the post-survey screening mechanism, we obtained 173 valid responses. The demographics of these valid respondents are showcased in \hyperref[tab:demographics]{Table \ref{tab:demographics}}, which highlights the diversity of the participant sample.

\begin{longtable}{p{0.2\textwidth}p{0.6\textwidth}}
    \caption{Demographic Summary of Survey Participants}
    \label{tab:demographics}\\
    \toprule
    Characteristic & Description \\
    \midrule
    \endfirsthead
    
    \caption[]{Demographic Summary of Survey Participants (continued)}\\
    \toprule
    Characteristic & Description \\
    \midrule
    \endhead
    
    \bottomrule
    \endfoot
    
    \bottomrule
    \endlastfoot
    
    \textbf{Gender} & Male (N = 119), Female (N = 54), Others (N = 0) \\
    \textbf{Age} & Minimum: 18, Maximum: 68, Mean: 27, SD: 7.748; Age distribution: $<$23 (25\%), 23--32 (50\%), $>$32 (25\%) \\
    \textbf{Country of Residence} & South Africa (N = 52), United Kingdom (N = 18), Canada (N = 16), Portugal (N = 16), Poland (N = 10) \\
    \textbf{Country of Birth} & South Africa (N = 46), United Kingdom (N = 14), Portugal (N = 14), Poland (N = 10), Canada (N = 8) \\
    \textbf{Employment Status} & Employed full-time (N = 91), Employed part-time (N = 21), Unemployed (N = 31), Other (N = 30) \\
    \textbf{Occupation} &  Professional specializing in IT (N = 80, 46\%), Student specializing in IT-related fields (N = 53, 30.5\%), Others (N = 41, 23.6\%) \\
    \textbf{Years of Programming Experience} & Mean = 4.13 years, Standard deviation = 4.365; 25th percentile: 2 years, 50th percentile: 3 years, 75th percentile: 5 years \\
    \textbf{Years of JavaScript Experience} & Mean = 4.13 years, Standard deviation = 4.365; 25th percentile: 1 year, 50th percentile: 1.5 years, 75th percentile: 3 years \\
\end{longtable}

\section{Results}
\subsection{Preliminary Analysis}
Before testing the hypotheses, we first calculated the latent variables ($B_1$) \textit{development expertise} and ($Y$) \textit{latent code similarity}. 

\subsubsection{Development Expertise}
\label{sec:Development Expertise}
We created the latent variable \textit{development expertise ($A_3$)} in two steps. Firstly, we used Cronbach's alpha \cite{128} to test whether all the questions used for assessing programmer factors were measuring the same underlying construct. Secondly, we used principal component analysis (PCA) \cite{120} to create the latent variable ($B_1$) \textit{development expertise} with observed variables $B_{11a}$, $B_{11b}$, $B_{12a}$, $B_{12b}$, $B_{13a}$, and $B_{13b}$. 

Cronbach's alpha is a measure of reliability that determines the internal consistency of multiple items designed to measure the same construct \cite{128}. The value of Cronbach's alpha of $B_{11a}$, $B_{11b}$, $B_{12a}$, $B_{12b}$, $B_{13a}$, and $B_{13b}$ was 0.833, higher than the cut-off value 0.8 \cite{128}. This result indicated a strong internal consistency among all the scales of observed programmer factors. 

We then performed PCA on $B_{11a}$, $B_{11b}$, $B_{12a}$, $B_{12b}$, $B_{13a}$, and $B_{13b}$ to compute ($A_3$) \textit{development expertise}. We tested the linearity asssumption and sampling adequacy assumption \cite{123} to ensure that the collected data were suitable for PCA. We have tested the Kaiser-Meyer-Olkin (KMO) measure \cite{123} and Bartlett's test of sphericity \cite{124} to ensure the sampling adequacy assumption is not violated. The detailed information of KMO measure, Bartlett's test of sphericity, and other detail of PCA analysis is shown in \hyperref[appendix:PCA1]{\ref{appendix:PCA1}}. We then used the scree plot criterion \cite{121} to choose latent factors that represent the observed variables. The scree plot criterion \cite{121} suggested that only the first factor should be chosen. Keeping the first component explained 54.729\% of the variance, which was acceptable \cite{121}. The component matrix is shown in \hyperref[table:component]{Table \ref{table:component}}, indicating that the chosen factor could cover all observed variables. The interpretation of the data was consistent with the concepts of development expertise, since it involves positive coefficients for overall programming experience, programming experience in particular programming languages, and task-specified knowledge familiarity.


\subsubsection{Latent Code Similarity}
\label{sec:Latent Code Similarity}
To simplify the data analysis and see the big picture of code similarity, we used PCA to generate the independent variable \textit{latent code similarity} inferred from the four code similarity metrics listed in \hyperref[tab:similarity-metrics]{Table \ref{tab:similarity-metrics}}. The procedure of computing latent code similarity was the same as computing development expertise. One latent factor was selected to represent the \textit{TP}, \textit{TA}, \textit{LP}, and \textit{LA}. The retained factor explained 82.550\% of the total variances. As shown in \hyperref[table:component-y]{Table \ref{table:component-y}}, the retained latent variable could capture all the features of the four different code similarity metrics. The detailed PCA analysis of code similarity are shown in \hyperref[appendix:PCA2]{\ref{appendix:PCA2}}.


To validate the suitability of code similarity as a proxy for AI code adoption, we randomly selected 20 pairs of participants’ code—specifically, their original submissions before encountering the AI solution and their revised submissions after reviewing the AI-generated code. For each pair, we systematically evaluated two metrics: the similarity between their revised code and the AI solution, and the actual extent of AI code adoption.

Adoption levels were categorized as follows: 

\textbf{Full adoption}: The participant entirely discarded their original code and adopted the AI-generated solution in its entirety, with no remnants of their initial implementation retained.

\textbf{Partial adoption}: The participant preserved portions of their original code while incorporating specific components of the AI-generated solution without full replication. The specific components adopted by participants are clarified in the \textit{Details} column in \hyperref[table:adoption-details]{Table \ref{table:adoption-details}}. Examples include: additions of defensive programming elements (to enhance code robustness, such as incorporating \texttt{if (!head) return head} from the AI-generated solution to prevent null pointer errors), and adjustments to control structures (e.g., switching from recursive to iterative implementations to align with the AI solution).

\textbf{No evidence of adoption}: The revised code contains no recognizable elements derived from the AI solution. 

As shown in \hyperref[table:adoption-details]{Table \ref{table:adoption-details}}, our analysis revealed a clear hierarchical pattern: revised code classified as full adoption showed significantly higher similarity to the AI solution than that with partial adoption, while partial adoption cases consistently exhibited greater similarity than those with no adoption. This ordered relationship confirms that code similarity effectively reflects the degree of AI code adoption, thereby validating its use as a reliable proxy measure.

\begin{ThreePartTable}
\begin{longtable}{L{0.1\columnwidth} L{0.15\columnwidth} L{0.1\columnwidth} L{0.28\columnwidth} L{0.22\columnwidth}}
\caption{Adoption details of programming tasks}
\label{table:adoption-details}\\
\toprule
LeetCode Number\tnote{a} & Participant's Prolific ID\tnote{b} & Similarity\tnote{c} & Adoption Level & Details \\
\midrule
\endfirsthead 

\caption{Adoption details of programming tasks (continued)}\\
\toprule
LeetCode Number & Participant's Prolific ID & Similarity & Adoption Level & Details \\
\midrule
\endhead 

\bottomrule
\endfoot 

\bottomrule
\endlastfoot 

83 & 6511 & 0.67 & partial adoption & defensive programming \\
83 & 60cd & 0.81 & full adoption &  \\
83 & 5f20 & 0.95 & full adoption &  \\
83 & 5f0d & -0.01 & partial adoption & defensive programming, control structure \\
98 & 657d & 0.95 & full adoption &  \\
98 & 6525 & -0.11 & partial adoption & defensive programming \\
98 & 650b & 0.95 & full adoption &  \\
98 & 60ae & -1.12 & no evidence of adoption &  \\
1823 & 6611 & 0.95 & full adoption &  \\
1823 & 65e1 & -1.21 & no evidence of adoption &  \\
1823 & 64eb & 0.95 & full adoption &  \\
1823 & 6129 & -0.03 & partial adoption & defensive programming \\
32 & 654c & 0.95 & full adoption &  \\
32 & 627b & 0.95 & full adoption &  \\
32 & 5df6 & -0.15 & partial adoption & defensive programming, control structure \\
32 & 5ba0 & 0.95 & full adoption &  \\
383 & 6558 & -1.74 & no evidence of adoption &  \\
383 & 6122 & 0.95 & full adoption &  \\
383 & 659e & 0.95 & full adoption &  \\
383 & 6609 & -1.78 & no evidence of adoption &  \\

\end{longtable}

\begin{tablenotes}
\item[a] LeetCode Number is a unique identifier assigned to each problem on the LeetCode platform for easy reference and search.
\item[b] The \textit{Participant’s Prolific ID} column displays the first 4 digits of each participant’s Prolific ID. This ID serves as the unique identifier for participants on the Prolific platform.
\item[c] Negative values in the \textit{Similarity} column exist because data was normalized.
\end{tablenotes}
\end{ThreePartTable}

\subsection{Hypotheses Testing}
\label{subsec:Hypotheses Testing}
We employed linear regression to examine the influence of ($A_{1}$) \textit{presence of comments} and ($B_{1}$) \textit{development expertise} on (Y) \textit{latent code similarity}, using a data analysis approach consistent with previous studies \cite{32, 57}. We hypothesize that the level of development expertise exerts a moderation effect \cite{160} on the influence of comments on programmers' adoption of AI-generated code because more experienced programmers may rely less on comments to understand and integrate the code, while less experienced programmers may find comments crucial for comprehension and adoption. To explore the potential moderation effect, we added an interaction term into the linear regression models. The interaction term was calculated as the product of ($A_{1}$) \textit{presence of comments} and ($B_{1}$) \textit{development expertise}, denoted as $A_1 \times B_1$. 

We firstly included ($A_{1}$) \textit{presence of comments}, ($B_{1}$) \textit{development expertise} and the interaction term $A_1 \times B_1$ into the linear regression. The overall model significantly predicted ($Y$) \textit{latent code similarity}, $F(3, 173) = 2.724, p=0.046$. The coefficient table for the factors $A_{1}$, $B_{1}$, and $A_1 \times B_1$ is shown in \hyperref[tab:lr0]{Table \ref{tab:lr0}}. The results demonstrated that ($A_1$) the presence of comments could significantly increase programmers' adoption of AI-generated code ($\beta = 0.336, p = 0.027$), supporting hypothesis \textbf{$H_1$}\footnote{To address task-specific variability and uneven dropouts across the five programming tasks, we re-analyzed the data using a mixed-effects model, which included a random intercept for task as the random effect and retained the presence of comments as the fixed effect. The re-analysis confirmed our core findings: the presence of comments remained a significant predictor of AI code adoption (\(p = 0.019\)), supporting \(H_1\)}. We anecdotally observed that comments were copied in the majority of cases and that participants also tended to copy function signatures. A systematic analysis of these behaviors is part of our future work.

As showcased in \hyperref[tab:lr0]{Table \ref{tab:lr0}}, neither ($B_{1}$) \textit{development expertise} nor the interaction term between ($B_{1}$) \textit{development expertise} and ($A_{1}$) \textit{presence of comments} had a significant influence on code adoption ($p>0.05$). This result indicated that the level of development expertise did not moderate the influence of ($A_{1}$) \textit{presence of comments}\footnote{To further validate robustness, we re-estimated the model using years of programming experience (highest-loading indicator of development expertise) instead of the composite variable $B_{1}$. Results confirmed stability: presence of comments remained significant (\( p = 0.037 \)); other variables remained insignificant (programming experience: \( p = 0.801 \); interaction term: \( p = 0.561 \)).}. Programmers tended to adopt more AI-generated code when comments were present, regardless of their development expertise. The insignificance of $B_{1}$ and $A_1 \times B_1$ suggested that ($B_{1}$) \textit{development expertise} did not diminish the positive effect of ($A_{1}$) \textit{presence of comments} on programmers' adoption of AI-generated code. Therefore, we reject hypothesis \textbf{$H_2$}.  

\begin{table}[htb]
\centering
\begin{threeparttable}
\caption{Coefficients and Significance of Key Factors}
\label{tab:lr0}
\begin{tabular}{cccc}
\toprule 
Variable & Coefficient & Standardized Coefficient & Significance  \\
\midrule
($A_{1}$) presence of comments & 0.336 & 0.168 & 0.027  \\
($B_{1}$) development expertise & -0.113 & -0.113 & 0.253  \\
($A_1 \times B_1$) interaction term & -0.015 & -0.010 & 0.922  \\
\bottomrule 
\end{tabular}
\end{threeparttable}
\end{table}

To further confirm and validate the results in \hyperref[tab:lr0]{Table \ref{tab:lr0}}, we added all control variables, including ($B_2$) \textit{gender}, ($B_{3}$) \textit{age}, ($B_{4f}$) \textit{fear of AI code generators}, and ($B_{4a}$) \textit{attitudinal acceptance of AI code generators}, into our initial linear regression model. The coefficients and significance of all variables included in the second model are presented in \hyperref[tab:lr1]{Table \ref{tab:lr1}}. \hyperref[tab:lr1]{Table \ref{tab:lr1}} indicated that ($A_{1}$) \textit{presence of comments} was a significant predictor of (Y) \textit{latent code similarity}, while ($B_{1}$) \textit{development expertise} and the interaction term $A_1 \times B_1$ remained insignificant. The second linear regression model confirmed the results of the initial regression model: even when accounting for all control variables, the presence of comments increases programmers' adoption of AI-generated code regardless of programmers' development expertise. \hyperref[tab:lr1]{Table \ref{tab:lr1}} also showed that all control variables were insignificant, including participants' attitudes towards AI code generators. The discrepancy between self-reported attitudes and actual behavior demonstrates the limitation of overly relying on self-reported intention of adopting AI coding tools, further highlighting the importance of adopting objective measurements of programmers' code adoption behavior.  

\begin{table}[htb]
\centering
\begin{threeparttable}
\caption{Coefficients and Significance of Key Factors and Control Variables}
\label{tab:lr1}
\begin{tabular}{@{}p{4cm} c c c@{}}
\toprule 
Variable & Coefficient & Standardized Coefficient & Significance \\
\midrule
($A_{1}$) presence of comments & 0.354 & 0.177 & 0.022  \\
($B_{1}$) development expertise & -0.138 & -0.138 & 0.176  \\
($A_1 \times B_1$) interaction term & -0.021 & -0.014 & 0.889  \\
($B_{2}$) gender & 0.117 & 0.008 & 0.490  \\
($B_{3}$) age & 0.009 & -0.084 & 0.401  \\
($B_{4f}$) fear of AI code generators & 0.003 & 0.054 & 0.921  \\
($B_{4a}$) attitudinal acceptance of AI code generators & -0.047 & 0.067 & 0.280  \\
\bottomrule 
\end{tabular}
\end{threeparttable}
\end{table}

\subsection{Post-hoc Analysis}
While the latent code similarity variable captures the overall effects of comments and development expertise, we conducted post-hoc analyses to examine their more nuanced effects on individual code similarity metrics—specifically, line-level and token-sequence-level similarities. We re-ran the linear regression models with \textit{TA}, \textit{TP}, \textit{LA}, and \textit{LP} as dependent variables separately. The results of these regression models are presented in \hyperref[appendix:regression]{\ref{appendix:regression}}.

These analyses show that for line-level similarities (\textit{LA} and \textit{LP}), the presence of comments (\textit{$A_{1}$}) is not a significant predictor, whereas development expertise (\textit{$B_{1}$}) is. In contrast, for token-sequence-level similarities (\textit{TA} and \textit{TP}), comment presence (\textit{$A_{1}$}) is significant, while development expertise (\textit{$B_{1}$}) is not. This suggests that the global effects observed in our primary model were primarily driven by token-sequence-level similarities.

As introduced in subsection~\ref{subsec:Dependent Variable}, line-level code similarities primarily reflect direct copy-paste behavior, whereas token-sequence-level similarities capture the adoption of structural patterns. A possible explanation for the differing predictors is that less experienced programmers may be more inclined to directly copy and paste lines from the AI-generated solution due to limited confidence or an inability to devise their own approach. In contrast, more experienced programmers may be able to construct a rough solution independently and thus adopt the structural logic of the AI-generated code rather than its exact lines. This behavioral difference may explain why development expertise predicts line-level similarity but not token-level similarity.

These findings underscore the importance of being cautious in the operationalization of “code similarity,” as different aspects of similarity may be predicted by different factors. Nevertheless, given our study’s focus on the broader effects of comments and expertise, we continue to interpret and discuss the latent code similarity variable in the main manuscript, as it captures multiple dimensions of code similarity.
\color{black} 

\section{Discussion and Implications}
Our research demonstrates that comments—an essential element in facilitating code comprehension—remain significant in shaping programmers' adoption of AI-generated code within the context of lightweight JavaScript function coding. The presence of comments significantly encourages programmers to adopt AI-generated code, regardless of their development expertise. Both novice and expert programmers tended to adopt more AI-generated code when it included comments. 

For novice programmers, the influence of comments on the adoption of AI-generated code can be understood through the Unified Theory of Acceptance and Use of Technology (UTAUT) \cite{158}. According to UTAUT, technologies that are perceived as easier to use are more likely to be adopted \cite{158}. Commented code is typically easier to comprehend, and this enhanced comprehensibility can strengthen users’ intention to adopt AI-generated solutions.

Previous studies have demonstrated the profound influence of comments on code comprehension \cite{40, 57}. Kolodziej discovered that code snippets with comments significantly decreased the time needed for code comprehension and improved the correctness rates in code comprehension tests \cite{57}. B{\"o}rstler and Paech evaluated the influence of comment quality on code comprehension, and found that even comments of poor quality can lead to a higher perceived code comprehension compared to code snippets without comments \cite{40}. 

The comprehensibility of AI-generated code plays an important role in participants' adoption decisions. Prior studies have shown that many users perceive code understandability as a limitation of AI coding tools \cite{5, 9, 81}. GitHub Copilot user studies discovered that users found it difficult to understand the code generated by Copilot \cite{5, 9}, which led some to refrain from using it for programming tasks \cite{5}. In a large-scale survey, Liang et al. found that 16\% of participants rated "I don't understand the code written by code generation tools" as "Very important" or "Important" when asked about reasons for not using AI coding tools \cite{81}. These findings align with UTAUT’s emphasis on effort expectancy—users are less likely to adopt a technology they perceive as difficult to use \cite{158}. When AI-generated code lacks comprehensibility, it increases cognitive load and reduces perceived ease of use, thereby discouraging adoption.

Our study revealed that the presence of comments can also increase the adoption of AI-generated code for expert programmers. Previous studies have discovered that expert programmers were more capable of understanding source code \cite{32, 45}. It is reasonable to extrapolate that expert programmers were more capable of understanding the AI-generated code without comments. However, they were still more likely to adopt AI-generated code that was commented. This indicates that the role of comments extends beyond mere comprehensibility. For expert programmers, the presence of comments may serve as a signal of higher code quality—an aspect they prioritize more strongly due to their elevated expectations and professional standards. Code comments could be leveraged to improve the readability, maintainability, and reliability of software \cite{161}. Expert programmers are expected to produce high-quality code that exhibits several desirable characteristics, including enhanced readability and good maintainability \cite{116}. Commented code was more capable of meeting the requirements of expert programmers. In this case, their adoption may be more driven by performance expectancy in UTAUT \cite{158}, specifically the belief that using well-documented AI-generated code will enhance their capability to execute high-quality, maintainable development work.

To encourage more programmers to utilize AI coding tools and enhance the adoption rates of AI-generated code, the developer of AI coding tools should consider ensuring the presence of comments in generated code snippets and increasing the quality of accompanying documentation. 

\section{Limitations and Threats to Validity}
\textbf{Construct Validity}. The measurement of development expertise might not accurately reflect the true level of participants' development expertise. Development expertise was quantified as a linear combination of ($A_{31}$) \textit{programming experience}, ($A_{32}$) \textit{programming language proficiency}, and ($A_{33}$) \textit{domain knowledge familiarity}. Although we incorporated ostensibly objective measures, such as the number of years participants had focused primarily on programming in their work or studies, these metrics still relied on self-reporting, which can be inherently subjective and potentially unreliable. Baltes and Diehl investigated the validity of self-assessed expertise ratings and years of programming experience among various groups of programmers \cite{116}. They concluded that neither the self-evaluated experience ratings nor the years of experience could reliably measure programmers' actual expertise \cite{116}.  

Measuring AI-generated code adoption through similarity metrics can be objective but may miss some aspects of adoption. While our metrics capture the adoption of code structures and exact lines of code, they overlook the adoption of stylistic aspects, such as identifier naming styles and function declaration styles. Specifically, \textit{TP} and \textit{TA} ignore identifier name changes, and \textit{LP} and \textit{LA} assess only exact line matches. Thus, if a participant adopts naming styles of the AI-generated code, like the camel case naming style, our metrics will not capture this. Additionally, our metrics also overlook cases where programmers may adopt the function declaration style used in AI-generated code, such as favoring arrow functions over traditional function expressions when working with JavaScript.

Another potential constraint on using code similarity as a measure of code adoption is the risk of chance overlap—defined as coincidental similarities between the AI-generated solution and a participant’s code that emerge not from deliberate adoption of the AI output, but because the participant independently developed a solution resembling the AI’s. While our analysis of 20 randomly selected code pairs (comparing solutions before and after exposure to the AI) suggested that chance overlap generally does not undermine the metric’s validity, residual possibilities remain: in rare cases, such unintended coincidences could potentially compromise the metric’s accuracy in reflecting actual adoption.

\textbf{External Validity}. This study faces several threats to external validity. Firstly, the programming language was limited to JavaScript. We chose JavaScript for practical reasons, as using multiple languages would have required significantly more participants and resources. Previous studies have shown that the quality of AI-generated code varies across different languages \cite{80}. Limiting the study to only one programming language may not accurately capture the overall characteristics of AI-generated code in other languages. Future research should explore how AI-generated comments influence adoption across multiple programming languages, considering that different languages have varying conventions for code readability, documentation practices, and idiomatic structures.  

Secondly, the scope of the programming tasks was narrow and restricted to data structure and algorithm problems from LeetCode. We selected LeetCode for feasibility, as it is widely used in related studies \cite{63, 64, 80, 49, 82} and provides a familiar, controlled platform for experimentation. Although these problems may not fully represent real-world tasks, they allowed us to cover a range of difficulty levels and topics. This restriction may not accurately represent real-world software engineering contexts, such as web development, system design, and database integration. Additionally, the difficulty of the tasks could influence programmers’ adoption of AI-generated code. While we included problems of varying difficulty, we did not analyze whether programmers were more likely to adopt AI-generated code for easier or harder problems. It is possible that programmers struggling with complex problems may rely more heavily on AI-generated solutions, while those solving simpler problems may feel more confident in their own implementations. Future research should systematically explore how task complexity affects AI-generated code adoption.  

Thirdly, the complexity of the programming tasks was constrained, with a limited number of possible solutions and short solution lengths, which may not capture the open-ended nature of real-life software development tasks. In professional settings, developers often work on projects with multiple valid approaches, varying levels of abstraction, and domain-specific constraints. Further research could explore AI-generated code adoption in larger-scale software projects where contextual factors such as maintainability, codebase integration, and long-term readability play a role.  

Fourthly, the study used error-free AI-generated code, which did not reflect real-world scenarios where AI-generated code can contain errors and vulnerabilities \cite{140, 149}. In real-world applications, developers frequently need to evaluate AI-generated code for correctness and security before adopting it. Additionally, AI-generated code might contain subtle inefficiencies or biases in implementation that could impact long-term maintainability. Future studies could investigate how the presence of AI-generated errors influences adoption behavior, particularly whether programmers with higher expertise are more selective in integrating AI-generated solutions.  

Lastly, participants from Prolific are not necessarily representative of programmers, although we tried to mitigate this issue through pre-screening mechanisms. While our participants included those with full-time employment in IT and computer science students, the recruitment method may have introduced biases compared to recruiting professional software engineers working in the industry. Future research could replicate this study with industry practitioners to assess whether similar adoption patterns emerge in workplace settings.  

Future studies could enhance the generalizability of the results by expanding the scope and complexity of programming tasks, evaluating the use of erroneous AI-generated code, and recruiting a more diverse group of programmers through professional networks.  

\textbf{Internal Validity}. To maintain procedural control, we implemented methods to encourage participants to adhere to the designed procedures. One potential threat is related to the difficulty in preventing participants from copying solutions generated by external AI. To address this, participants were initially asked to solve each problem independently to motivate active engagement and filter out low-effort responses. This self-exclusion approach helped improve data quality by filtering out less meaningful responses.  

While acknowledging that external factors could have impacted the code produced, our primary focus remained on adoption behavior, which we believed to be relatively unaffected by these external influences. Even if participants used external AI tools to verify their work, they were still less likely to adopt code without comments, which stands as our central finding, underscoring the influence of comments on code adoption. However, external AI assistance could introduce biases, particularly if participants cross-referenced different AI-generated solutions before submitting their final code. Moreover, different AI tools may produce stylistically distinct solutions, and the AI model used in this study may exhibit implicit biases in generating code. Future studies should examine how different AI code generators influence programmers' adoption behaviors and whether developers prefer AI-generated solutions from specific models based on readability or correctness.  

Additionally, real-world software development is often constrained by deadlines, code review processes, and collaboration with other developers, all of which could influence code adoption. Our controlled environment did not fully replicate these constraints, and future research could examine how time pressure and team collaboration affect programmers’ willingness to adopt AI-generated code. Furthermore, it remains unclear whether repeated exposure to AI-generated solutions could lead to over-reliance on AI suggestions or whether programmers develop strategies to critically assess AI-generated code over time. Future studies could investigate the long-term effects of AI-assisted coding on software development practices.

\section{Conclusion and Future Work}
\label{sec:Conclusion and Contributions}

Our survey-based study explored how the presence of comments and development expertise influence programmers’ adoption of AI-generated JavaScript code. Participants reported their expertise, attempted to solve a programming problem independently, and then received AI-generated solutions with or without comments. Adoption was objectively measured through code similarity between participants' submissions and the AI-generated solutions. After data collection, we applied principal component analysis and linear regression to analyze the effects. 

Our study has made several important contributions. Firstly, we examined unexplored factors that could potentially affect programmers' adoption of AI-generated code. We found that programmers adopted more AI-generated JavaScript code when comments were present, regardless of development expertise. For industry practitioners, this finding has the potential to motivate developers of AI coding systems to place greater emphasis on annotating AI-generated code. For instance, future training in AI programming systems could incorporate more code with high-quality comments into the training set. This can help create more user-friendly AI coding tools, further boosting the productivity of AI coding assistants. For academia, this finding allows researchers to have a better understanding of the AI-human collaboration paradigm. Future studies could further expand this finding by evaluating what types of comments can help enhance programmers' adoption of AI-generated codes. 

Secondly, we proposed novel behavioral measurements for programmers' adoption of AI-generated code, which are valuable complements and alternatives to traditional self-reported data. Rather than directly inquiring of participants about the degree to which they adopted AI-generated code, we employed a behavioral approach. We measured the degree of code adoption by calculating the similarity between the participants' submitted code and the AI-generated code. We evaluated code similarity from two distinct dimensions: the entity perspective and the code similarity level. This new approach could provide behavioral and comprehensive insights about programmers' adoption of AI coding tools. We have also provided the replication package\footnote{The data preprocessing scripts, raw data, and the data analysis results can be found at: \url{https://github.com/ArthurLCW/CodeAdoptionResearch/tree/main/analysis}} to help future researchers utilize our measurements of code adoption. 

Thirdly, we developed a custom survey application that facilitated coding tasks\footnote{To access the survey app, please visit: \url{https://survey.changwen-software-engineering.xyz}; Source code of the survey app can be found at: \url{https://github.com/ArthurLCW/CodeAdoptionResearch/tree/main/survey-web}}. This survey application incorporated a code editor with advanced features, including syntax highlight and auto-completion, providing a coding experience similar to using VSCode \cite{163}. This allowed participants to complete coding tasks in a familiar environment. Future survey-based research can adapt and extend this application to suit their specific requirements, especially if the surveys include coding components. 

Future studies could build on our work by addressing its limitations and further validating our findings. One potential direction is adapting our survey application to support additional programming languages. Our app currently integrates the Monaco Editor—the same editor used in VSCode—which provides advanced features like syntax highlighting natively for JavaScript and TypeScript. To offer similar support for other languages, such as Python and Swift, future researchers would need to integrate the Language Server Protocol (LSP, a standardized interface that enables code editors to communicate with language-specific servers, providing consistent advanced features like auto-completion and syntax checking\footnote{More information regarding LSP: \url{https://microsoft.github.io/language-server-protocol/}}) with the Monaco Editor.

Additionally, future studies could expand our survey application to better reflect real-life programming scenarios. In our current study, participants are provided with pre-generated AI solutions, based on the assumption that they would typically query AI only once for a LeetCode problem. However, in real-world settings, developers often interact with AI tools iteratively, generating multiple solutions using different prompts. To more accurately simulate these interactions, future researchers should enhance the survey app to support on-demand generation of AI solutions at runtime, with the option to include or exclude comments depending on the experimental design.

Future researchers could extend our survey application to collect and analyze richer behavioral data, such as the number of copy and paste operations, to gain deeper insights into how programmers adopt AI-generated code. Specifically, the application could be enhanced to log user interactions, including copy actions, paste actions, and selection changes, to better capture participants' behavior during code adoption.

In addition to further validating our study results, future research could deepen our understanding of how comments influence programmers' adoption of AI-generated code. While there is ample prior research on the impact of comments on code comprehension \cite{37, 38, 39, 40, 161}, future studies could build on these designs to investigate how varying types, lengths, and qualities of comments affect the extent to which programmers adopt AI-generated code.

\newpage
\appendix
\begingroup  
\section{PCA Results for Development Expertise}
\label{appendix:PCA1}

We examined the linearity assumption \cite{121} through the correlation matrix in \hyperref[table:correlation_matrix]{Table \ref{table:correlation_matrix}}. As presented in \hyperref[table:correlation_matrix]{Table \ref{table:correlation_matrix}}, all observed variables have at least one correlation with another observed variable greater than 0.3. The correlation coefficient is widely used as an indicator of linearity, where the correlation coefficient larger than 0.3 indicates linearity \cite{122}. The matrix table indicated that every variable had a linear relationship with some other variables.

\begin{table}[ht]
\centering
\caption{Correlation matrix of observed developer factors}
\label{table:correlation_matrix} 
\begin{tabular}{@{}llclccc@{}} 
\toprule 
&  \textbf{$B_{11\text{sub}}$} 
&\textbf{$B_{11\text{obj}}$} &  \textbf{$B_{12\text{sub}}$} 
&\textbf{$B_{12\text{obj}}$} & \textbf{$B_{13\text{sub}}$} & \textbf{$B_{13\text{obj}}$} \\
\midrule 
\textbf{$B_{11\text{sub}}$} & 1.000 & 0.571 & 0.445 & 0.470 & 0.547 & 0.527 \\
\textbf{$B_{11\text{obj}}$} & 0.571 & 1.000 & 0.288 & 0.748 & 0.372 & 0.337 \\
\textbf{$B_{12\text{sub}}$} & 0.445 & 0.288 & 1.000 & 0.503 & 0.347 & 0.323 \\
\textbf{$B_{12\text{obj}}$} & 0.470 & 0.748 & 0.503 & 1.000 & 0.318 & 0.290 \\
\textbf{$B_{13\text{sub}}$} & 0.547 & 0.372 & 0.347 & 0.318 & 1.000 & 0.724 \\
\textbf{$B_{13\text{obj}}$} & 0.527 & 0.337 &  0.323 &0.290 & 0.724 & 1.000 \\
\bottomrule 
\end{tabular}
\end{table}

We tested the sampling adequacy assumption using the Kaiser-Meyer-Olkin (KMO) measure and Bartlett’s test of sphericity. The overall KMO value was 0.703, exceeding the 0.5 threshold \cite{123}, with all individual KMO values also above 0.5 (see \hyperref[table:KMO Measure]{Table \ref{table:KMO Measure}}), suggesting sufficient shared variance for factor analysis. Bartlett’s test of sphericity was significant ($p < 0.001$) \cite{124}, indicating that the correlation matrix was not an identity matrix. Together, these results confirmed the data’s suitability for factor analysis.

\begin{table}[ht]
\centering
\caption{KMO measures of observed developer factors}
\label{table:KMO Measure} 
\begin{tabular}{@{}cc@{}} 
\toprule
\textbf{Observed Variable} & \textbf{KMO Measure} \\
\midrule
\textbf{$B_{11\text{sub}}$} & 0.824 \\
\textbf{$B_{11\text{obj}}$} & 0.632 \\
\textbf{$B_{12\text{sub}}$} & 0.669 \\
\textbf{$B_{12\text{obj}}$} & 0.636 \\
\textbf{$B_{13\text{sub}}$} & 0.742 \\
\textbf{$B_{13\text{obj}}$} & 0.734 \\
\bottomrule 
\end{tabular}
\end{table}

PCA revealed two components with eigenvalues greater than 1, explaining 54.729\% and 18.169\% of the total variance, respectively. Based on the scree plot criterion \cite{121,125}, we retained only the first component, which appears before the inflection point in the scree plot (\hyperref[fig:scree-plot]{Fig. \ref{fig:scree-plot}}). This decision was supported by the high explained variance (54.729\%) and the interpretability of the component. We applied Varimax rotation with Kaiser normalization \cite{126,127}, and the rotation matrix (\hyperref[table:rotation]{Table \ref{table:rotation}}) showed that the first component adequately represented all observed variables. We then reran PCA to extract a single component and recalculated the loadings (\hyperref[table:component]{Table \ref{table:component}}). The resulting component aligned with the concept of Development Expertise ($B_1$), as it positively loaded on general programming experience, language proficiency, and domain-specific knowledge.

\begin{figure}[htbp]
    \centering
    \includegraphics[width=0.75\textwidth]{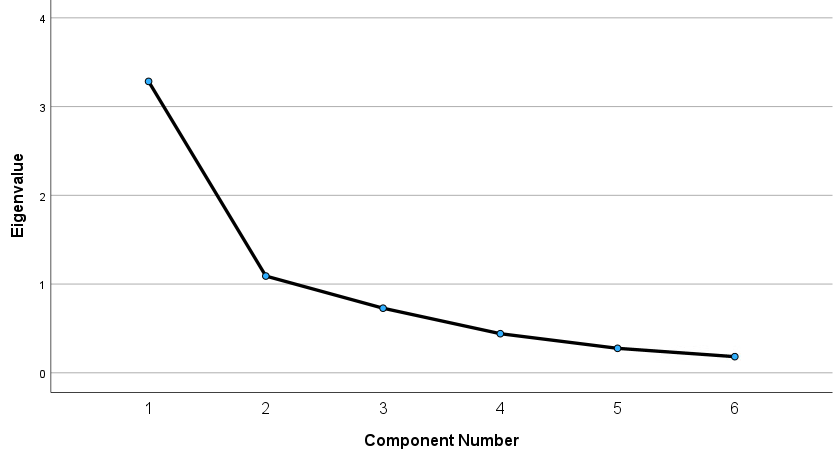}
    \caption{Scree plot of principal components for observed programmer factors}
    \label{fig:scree-plot}
\end{figure}

\begin{table}[htbp]
\centering
\caption{Rotation matrix with component loadings}
\label{table:rotation} 
\begin{tabular}{ccc}
\toprule
\textbf{Observed Variable} & \textbf{Component 1} & \textbf{Component 2} \\
\midrule
\texttt{$B_{11\text{sub}}$} & 0.572 & 0.315 \\
\texttt{$B_{11\text{obj}}$} & 0.922 & 0.103 \\
\texttt{$B_{12\text{sub}}$} & 0.158 & 0.898 \\
\texttt{$B_{12\text{obj}}$} & 0.853 & 0.189 \\
\texttt{$B_{13\text{sub}}$} & 0.205 & 0.886 \\
\texttt{$B_{13\text{obj}}$} & 0.564 & 0.589 \\
\bottomrule
\end{tabular}
\end{table}

\begin{table}[htbp]
\centering
\caption{Component matrix for development expertise}
\label{table:component} 
\begin{tabular}{cc}
\toprule
\textbf{Observed Variable} & \textbf{Coefficient} \\
\midrule
\texttt{$B_{11\text{sub}}$} & 0.751 \\
\texttt{$B_{11\text{obj}}$} & 0.813 \\
\texttt{$B_{12\text{sub}}$} & 0.748 \\
\texttt{$B_{12\text{obj}}$} & 0.758 \\
\texttt{$B_{13\text{sub}}$} & 0.722 \\
\texttt{$B_{13\text{obj}}$} & 0.635 \\
\bottomrule
\end{tabular}
\end{table}

Table \ref{table:communality} reports the communalities for each observed variable included in the PCA. These values represent the proportion of variance in each variable accounted for by the extracted principal component \cite{168}. Most variables show high communalities (e.g., \textit{$B_{12\text{obj}}$} = 0.861; \textit{$B_{13\text{sub}}$} = 0.827), indicating they are well explained by the component. This supports the validity of summarizing these variables into a single latent factor representing development expertise (\textit{$B_1$}).

\begin{table}[htbp]
\centering
\caption{Communalities for observed variables in development expertise}
\label{table:communality}
\begin{tabular}{cc}
\toprule
\textbf{Observed Variable} & \textbf{Communality} \\
\midrule
\texttt{$B_{11\text{sub}}$} & 0.665 \\
\texttt{$B_{11\text{obj}}$} & 0.763 \\
\texttt{$B_{12\text{sub}}$} & 0.427 \\
\texttt{$B_{12\text{obj}}$} & 0.861 \\
\texttt{$B_{13\text{sub}}$} & 0.827 \\
\texttt{$B_{13\text{obj}}$} & 0.832 \\
\bottomrule
\end{tabular}
\end{table}

\section{PCA Results for Code Similarity}
\label{appendix:PCA2}
Similar to the process for generating \textit{development expertise} ($B_1$), we first tested the linearity and sampling adequacy assumptions. Linearity was assessed using the correlation matrix in \hyperref[table:correlation_matrix_y]{Table \ref{table:correlation_matrix_y}}. All code similarity metrics showed at least one correlation above 0.6, with several exceeding 0.9 (e.g., between \textit{TP} and \textit{TA}; \textit{LP} and \textit{LA}), indicating strong inter-correlations \cite{122}. These high correlations satisfy the linearity assumption and suggest the metrics are closely related. The strong inter-correlations may be attributed to the short length of the code solutions (typically under 25 lines), which limits structural variability and amplifies similarity across metrics.

\begin{table}[ht]
\centering
\caption{Correlation matrix of code similarity metrics}
\label{table:correlation_matrix_y} 
\begin{tabular}{@{}lcccc@{}} 
\toprule 
& \textbf{TP} & \textbf{TA} & \textbf{LP} & \textbf{LA} \\
\midrule 
\textbf{TP} & 1.000 & 0.962 & 0.669 & 0.628 \\
\textbf{TA} & 0.962 & 1.000 & 0.687 & 0.688 \\
\textbf{LP} & 0.669 & 0.687 & 1.000 & 0.969 \\
\textbf{LA} & 0.628 & 0.688 & 0.969 & 1.000 \\
\bottomrule 
\end{tabular}
\end{table}

After testing linearity, we assessed sampling adequacy. The overall Kaiser-Meyer-Olkin (KMO) measure was 0.527, with all individual KMO values above 0.5 (\hyperref[table:KMO Measure y]{Table \ref{table:KMO Measure y}}), indicating sufficient shared variance for dimensionality reduction \cite{123}. Bartlett's test of sphericity was significant ($p < 0.001$), suggesting the correlation matrix was not an identity matrix \cite{124}. Together, these results confirmed that the sampling adequacy assumption was met.

\begin{table}[ht]
\centering
\caption{KMO measures of code similarity metrics}
\label{table:KMO Measure y} 
\begin{tabular}{@{}cc@{}} 
\toprule
\textbf{Code Similarity Metric} & \textbf{KMO Measure} \\
\midrule
\textbf{TP} & 0.515 \\
\textbf{TA} & 0.538 \\
\textbf{LP} & 0.537 \\
\textbf{LA} & 0.516 \\
\bottomrule 
\end{tabular}
\end{table}

PCA revealed one principal component with an eigenvalue above 1, explaining 82.550\% of the total variance. The scree plot (\hyperref[fig:scree-plot-y]{Fig. \ref{fig:scree-plot-y}}) confirmed that only the first component should be retained, as the second point marked the inflection point. This component captured the core variance across all four code similarity metrics and was interpretable as a unified latent variable. We used it as the dependent variable, with its component loadings shown in \hyperref[table:component-y]{Table \ref{table:component-y}}.

\begin{figure}[htbp]
    \centering
    \includegraphics[width=0.75\textwidth]{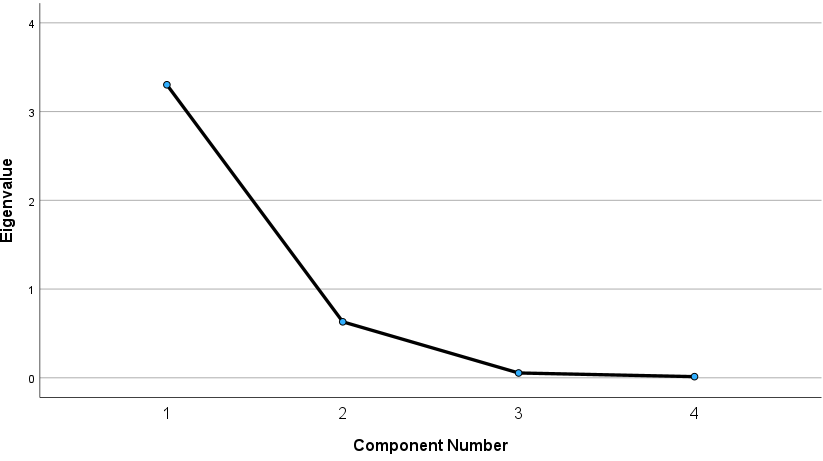}
    \caption{Scree plot of principal components for code similarity metrics}
    \label{fig:scree-plot-y}
\end{figure}

\begin{table}[ht]
\centering
\caption{Component matrix for latent code similarity}
\label{table:component-y} 
\begin{tabular}{cc}
\toprule
\textbf{Code Similarity Metric} & \textbf{Coefficient} \\
\midrule
\texttt{TP} & 0.915 \\
\texttt{TA} & 0.905 \\
\texttt{LP} & 0.896 \\
\texttt{LA} & 0.918 \\
\bottomrule
\end{tabular}
\end{table}

We examined the communalities to assess how much of each code similarity metric's variance is captured by the extracted principal component. As shown in \hyperref[table:communalities-y]{Table \ref{table:communalities-y}}, all four code similarity metrics had high extraction values, ranging from 0.803 to 0.843. This indicates that the retained component effectively represents the shared variance across these metrics, supporting the dimensionality reduction.

\begin{table}[ht]
\centering
\caption{Communalities for code similarity metrics}
\label{table:communalities-y}
\begin{tabular}{cc}
\toprule
\textbf{Code Similarity Metric} & \textbf{Extraction} \\
\midrule
\texttt{TP} & 0.838 \\
\texttt{TA} & 0.818 \\
\texttt{LP} & 0.803 \\
\texttt{LA} & 0.843 \\
\bottomrule
\end{tabular}
\end{table}

\section{Post-Hoc Regression Analyses on Individual Code Similarity Metrics}
\label{appendix:regression}
We conducted post-hoc regression analyses to examine how development expertise (\textit{$B_{1}$}) and the presence of comments (\textit{$A_{1}$}) influence individual code similarity metrics. Specifically, we analyzed: token-sequence-level similarity from the participant’s perspective (\textit{TP}), token-sequence-level similarity from the AI’s perspective (\textit{TA}), line-level similarity from the participant’s perspective (\textit{LP}), and line-level similarity from the AI’s perspective (\textit{LA}). 

We employed the same linear regression model described in Subsection~\ref{subsec:Hypotheses Testing}, replacing the latent code similarity variable with each of the four individual similarity metrics as dependent variables. The regression results are summarized in Tables~\ref{tab:lr_ta} through~\ref{tab:lr_lp}.

\begin{table}[htb]
\centering
\begin{threeparttable}
\caption{Regression Results with \textit{TA} as Dependent Variable}
\label{tab:lr_ta}
\begin{tabular}{cccc}
\toprule 
Variable & Coefficient & Standardized Coefficient & $p$-value \\
\midrule
($A_{1}$) Presence of comments & 0.157 & 0.208 & 0.006 \\
($B_{1}$) Development expertise & -0.012 & 0.012 & 0.835 \\
($A_{1} \times B_{1}$) Interaction term & -0.021 & 0.005 & 0.900 \\
\bottomrule 
\end{tabular}
\end{threeparttable}
\end{table}

\begin{table}[htb]
\centering
\begin{threeparttable}
\caption{Regression Results with \textit{TP} as Dependent Variable}
\label{tab:lr_tp}
\begin{tabular}{cccc}
\toprule 
Variable & Coefficient & Standardized Coefficient & $p$-value \\
\midrule
($A_{1}$) Presence of comments & 0.118 & 0.156 & 0.041 \\
($B_{1}$) Development expertise & -0.002 & -0.004 & 0.964 \\
($A_{1} \times B_{1}$) Interaction term & -0.001 & -0.003 & 0.980 \\
\bottomrule 
\end{tabular}
\end{threeparttable}
\end{table}

\begin{table}[htb]
\centering
\begin{threeparttable}
\caption{Regression Results with \textit{LA} as Dependent Variable}
\label{tab:lr_la}
\begin{tabular}{cccc}
\toprule 
Variable & Coefficient & Standardized Coefficient & $p$-value \\
\midrule
($A_{1}$) Presence of comments & 0.093 & 0.138 & 0.064 \\
($B_{1}$) Development expertise & -0.070 & -0.207 & 0.034 \\
($A_{1} \times B_{1}$) Interaction term & -0.011 & -0.020 & 0.833 \\
\bottomrule 
\end{tabular}
\end{threeparttable}
\end{table}

\begin{table}[htb]
\centering
\begin{threeparttable}
\caption{Regression Results with \textit{LP} as Dependent Variable}
\label{tab:lr_lp}
\begin{tabular}{cccc}
\toprule 
Variable & Coefficient & Standardized Coefficient & $p$-value \\
\midrule
($A_{1}$) Presence of comments & 0.070 & 0.109 & 0.147 \\
($B_{1}$) Development expertise & -0.068 & 0.004 & 0.032 \\
($A_{1} \times B_{1}$) Interaction term & -0.211 & 0.009 & 0.930 \\
\bottomrule 
\end{tabular}
\end{threeparttable}
\end{table}

\endgroup

\newpage
\bibliographystyle{elsarticle-num}

\end{sloppy}
\end{document}